\documentclass[aip,
               preprint,
               amsmath,
               superscriptaddress,
               floatfix,
               nobibnotes,
               ]{revtex4-2}

\usepackage{amssymb}
\usepackage{mathtools}
\usepackage[mathscr]{eucal}
\usepackage[dvips]{graphicx}
\usepackage{color}
\usepackage[utf8]{inputenc}
\usepackage[flushleft]{caption}
\usepackage{subfigure}
\usepackage{rotating}
\usepackage{txfonts}
\usepackage{booktabs}
\usepackage[unicode]{hyperref}
\usepackage{bm}
\usepackage{graphicscache}

\DeclareMathAlphabet{\mathpzc}{OT1}{pzc}{m}{it}

\def\figfoot{H + H$_2$ Dynamics}
\def\mathbi#1{\textbf{\em #1}}

\makeatletter

\makeatletter

\newcommand{\figcaption}[2]{
    \noindent {\bf Figure \ref{#1}:} #2
    \vspace{1cm}}
    
\begin{document}


\title{Uncertainty-Driven Self-Refinement of Adaptive Kolmogorov–Arnold
Networks for Fitting Nonadiabatic Potential Energy Surface}

\author{Jinke Yu}
\affiliation{Theoretical Chemistry, Department of Chemistry, 
             Northwestern Polytechnical University,
             West Youyi Road 127, 710072 Xi'an,
             China}

\author{Xingyu Zhang}
\affiliation{Theoretical Chemistry, Department of Chemistry, 
             Northwestern Polytechnical University,
             West Youyi Road 127, 710072 Xi'an,
             China}
             
\author{Qingyong Meng*}
\thanks{To whom correspondence should be addressed}
\email{qingyong.meng@nwpu.edu.cn}
\affiliation{Theoretical Chemistry, Department of Chemistry, 
             Northwestern Polytechnical University,
             West Youyi Road 127, 710072 Xi'an,
             China}

\date{\today}


\begin{abstract}

\noindent {\bf Abstract}:
In this work, we propose an uncertainty-driven self-refinement learning
method using adaptive Chebyshev-polynomial-based Kolmogorov–Arnold networks
(ChebyKANs) to build potential energy surface (PES). Unlike conventional
approaches, our method employs a main-model-plus-residual-model architecture
for automatic sampling based on a hybrid criterion, where sampling and
fitting are decoupled. To benchmark our method, the non-adiabatic two-state
PES of the H + H$_2$ system is constructed using a database computed at
the multireference configuration interaction (MRCI) level with a rather
large active space. Our calculations show that the proposed architectural
design makes the efficiency gain, and that KAN offers enhanced interpretability
over conventional neural network (NN) fitting methods through its mathematically
well-understood basis functions. Furthermore, the present calculations
accurately reproduce all important conformations on the PES, including
the transition state and the conical intersection. Then, the multi-configurational
time-dependent Hartree (MCTDH) method reproduces the non-adiabatic
dynamics with results in close agreement with previously reported findings.
{\it Active Learning}; {\it Uncertainty Driven}; {\it Self-Refinement}

\end{abstract}

\maketitle


\section{Introduction\label{sec:intro}}

Since the potential energy surface (PES) is a prerequisite of dynamics
calculations, various fitting approaches
\cite{eve05:134110,beh07:014705,beh07:146401,beh11:17930,beh15:1032,che15:1241,bar18:041048,kam18:241702,wan19:4513,fu19:145,gua18:823,son19:114302,son22:1983,son22:11128,son24:597}
were reported in the literature. In the last two decades, it was
proved \cite{att97:1069,sch07:253} that the neural networks (NNs),
which was originally designed for numerical simulation of the nervous
activity in brain, have a capability to approach any functional form
with arbitrary precision because of the universal approximation theorem.
Then the NN fitting approach
\cite{fu19:145,hed19:123,xu19:1151,lux18:23095,wan18:17960,gua17:224307,zha16:135,che15:1241,beh07:014705,son19:114302,son22:1983}
for constructing PES was proposed and fast developed. For example, the
NN fitting approach and its various improvements were successfully
implemented to construct the PESs
\cite{fu19:145,hed19:123,xu19:1151,lux18:23095,wan18:17960,gua17:224307,zha16:135,che15:1241,beh07:014705,son19:114302}.
To make the NN function comply with the atomic permutation symmetry,
Guo and Zhang and co-workers \cite{li13:204103,jia13:054112,jia14:034109}
employed the permutation invariant polynomial (PIP) as input in NN fitting,
where the original PIP approach was designed by Bowman and co-workers
\cite{bow11:8094,cza12:044307,li12:041103} to consider the permutation
invariant of the same atoms. Since the PIPs instead of the coordinates
are input neurons, the PIP-NN approach \cite{li13:204103,jia13:054112,jia14:034109}
can be further simplified by just taking some parts of the polynomials
with fundamental invariance (FI), yielding the FI-NN approach \cite{sha16:071101,pin18:26315,lux18:174303}.
On the other hand, committee NN models \cite{chr20:104105} was developed to
provide generalization error estimates, enabling active learning strategies.
By this concept, one can not only identify the most relevant configurations
to build up training set in an active learning procedure but also control
the generalization error \cite{chr20:104105}. By two different NN models
to actively locate the point where the PES is least confident, Jiang and
co-workers \cite{lin20:154104,lin21:2691} developed an uncertainty-driven
active learning strategy to automatically and efficiently generate
high-dimensional NN-based PESs. In addition, Gaussian process regression
(GPR) was found not only to be an efficient method for constructing
multidimensional PESs \cite{gua18:823,son19:114302,son22:1983,son22:11128,son24:597}
but also to provide a well-established model for active learning
\cite{ele18:17414,gua18:823,son19:114302,son22:1983}. This is not surprising
because GPR provides not only the posterior mean but also the posterior
variance, while GPR was proved to be equivalent to a single-layer NN
structure with infinite neurons \cite{son22:1983}.

Despite its advantages, NNs face notable limitations in terms of parameter
efficiency and interpretability, where the NNs often require a large
number of parameters to achieve high accuracy. This requires an amount
of {\it ab initio} energy data on the similar order of magnitude as the
number of NN parameters to construct high-dimensional PES. Moreover,
employing fixed activation functions, NNs are less adaptable to complex
data patterns without extensive post-analysis. Therefore, noting the
computational cost in extensive {\it ab initio} energy calculations, the
construction of high-dimensional PES becomes extremely difficult and
costly. Inspired by the Kolmogorov-Arnold representation theorem \cite{kol57:953},
Kolmogorov-Arnold Networks (KANs) \cite{liu25:041051} offer a novel approach by placing
learnable activation functions on the edges instead of the nodes. This
setup transforms each weight parameter into a learnable univariate
function, often parametrized as a spline. In this context, KANs sum
incoming signals at the nodes and apply nonlinear transformations
through these learnable functions \cite{liu25:041051}. This enhances flexibility and
adaptability of KANs. In this work, the Chebyshev-polynomial-based
Kolmogorov-Arnold network (ChebyKAN) is implemented to construct the
PES, which incorporates Chebyshev polynomials to parametrize the
learnable functions. Recognizing their capabilities of rapid convergence
and numerical stability, Chebyshev polynomials enable ChebyKAN to
improve the accuracy and efficiency in constructing the PES. Unlike
traditional NNs, ChebyKAN allows the network to adjust its response
to input data dynamically and effectively. This leads to high accuracy
with fewer parameters, offering a compact representation of high-dimensional
PES. Furthermore, we propose a uncertainty-quantification-driven active
sampling technique. Unlike previously reported active sampling techniques
\cite{ele18:17414,chr20:104105,lin20:154104,lin21:2691}, the present
sampling technique is implemented through a hybrid criterion that integrates NN
fitting errors, energy percentiles, and the standard deviation derived
from GPR. To test the present method for fitting and sampling, the non-adiabatic
two-state PES of the H + H$_2$ system is constructed through the database
computed at the multireference configuration interaction (MRCI) level
in conjunction with a rather large active space. Our numerical results
precisely locate the transition state and the conical intersection, which
are fully consistent with those reported in earlier studies. The
multi-configurational time-dependent Hartree product (MCTDH) method
\cite{mey90:73,man92:3199,mey98:3011,bec00:1,mey03:251,mey09:book,mey12:351}
is adopted to obtain the dynamics in close agreement with previously
reported findings.

The rest of this paper is organized as follows. Section \ref{sec:theore-work}
describes theoretical framework and computational details of the present work,
including the Hamiltonian model, regression and sampling methods, and wave
function propagation. Section \ref{sec:results} presents numerical results
and discussions of the present results. Section \ref{sec:con} concludes
with a summary.

\section{Theoretical Framework\label{sec:theore-work}}

Unless otherwise specified in this work, all conformation parameters,
bond length and angle, are given in {\AA} or degree, while energy is
given in hartree.

\subsection{ChebyKAN Approach\label{sec:kan-fitting}}

In this work, a Chebyshev polynomial-based Kolmogorov-Arnold network
(KAN), denoted by ChebyKAN, is employed for fitting the function terms
of the potential matrix, which is a continue function of nuclear
coordinates (see Figure S1 of the Supporting Information) in
a region of the configuration space (see Figure S2). The core idea of
the KAN model (see Figure S3) is to replace the fixed activation functions
on nodes in traditional neural networks (NNs) with learnable functions
on the edges, which are parametrized using Chebyshev polynomial bases.
This design enhances approximation efficiency and numerical stability
while maintaining expressive power of the model. In the Supporting
Infoimation, we give schematic illustration of the present ChebyKAN
architecture, where a two-hidden-layer configuration is adopted with
each hidden layer consisting of neurons and utilizing Chebyshev
polynomials of some degree. Given the input coordinates, the
Chebyshev polynomials in the first layer
are linearly combined to construct the input tensor for the first
hidden layer. The second layer follows the same logic, and the final
output tensor passes through the Chebyshev polynomials of the last
layer to yield the output value. Comparing with conventional MLP
architectures, KAN fundamentally interchanges the mathematical roles
of nodes and edges. Although a ChebyKAN with $l$ hidden layers effectively
contains $l+1$ polynomial layers, the polynomial parameterization depends
solely on a single parameter (the degree) which allows the model to maintain
a controlled parameter count while still achieving high accuracy.

In details, the $l$-th KAN layer transforms an input vector $\mathbi{x}_l$
to $\mathbi{x}_{l+1}=\boldsymbol{\Phi}_l(\mathbi{x}_l)$ by a matrix of
univariate functions $\Phi(\cdot)$. The complete $L$-layer network for
input vector $\mathbi{q}$ is a composition of $(\boldsymbol{\Phi}_{L-1}
\circ\cdots\circ\boldsymbol{\Phi}_0)\mathbi{q}$. In the present
implementation of ChebyKAN, the univariate function
$\Phi(\cdot)$ on an edge is parametrized as a linear combination of
Chebyshev polynomials $T_n(\cdot)=\cos(n\arccos(\cdot))$. For arbitrary
input variable $x$, the Chebyshev polynomials satisfy the recurrence
relations $T_{n+1}(x)=2xT_n(x)-T_{n-1}(x)$. Moreover, the roots of
$T_n(x)=0$ are given by $x_k=\cos((k+1/2)\pi/n)$, while the extrema
of $T_n(x)$, that is the solutions of $T_n(x)=\pm1$, are given by
$x_k=\cos(k\pi/n)$, where $k=0,1,\cdots,n$. These expressions are
useful in optimization of the KAN model. The forward pass of a layer
first normalizes the input features into the
domain $[-1,1]$ through hyperbolic tangent function. For each normalized
feature, Chebyshev polynomials up to a specified degree $K$ are evaluated
to form a polynomial basis tensor $\mathcal{T}\in\mathbb{R}^{\text{batch}
\times\text{in}\times(K+1)}$, where $\mathcal{T}_{b,i,n}=T_n(\cdot)$.
The final output is obtained by contracting this basis tensor with a
learnable coefficient tensor $\mathcal{C}\in\mathbb{R}^{\text{in}\times\text{out}\times(K+1)}$
through the summation,
\begin{equation}
y_{b,o}=\sum_{i=1}^{\text{in}}\sum_{n=0}^K\mathcal{C}_{i,o,n}\mathcal{T}_{b,i,n},
\label{eq:cheykan-nn-0}
\end{equation}
where $K$ is the highest degree of the Chebyshev polynomials. In this
work, the ChebyKAN models employs $K=6$ and $K=3$ for adiabatic potentials
and coupling terms, respectively, together with the NNs with $3$‑$128$‑$128$‑$1$
structure.

The present sampled conformations (see Figure S2) are distributed
in the ranges of $r_{\mathrm d}\in[0.1,5.5]$, $r_{\mathrm v}\in[0.2,3.6]$,
and $\theta\in[0.0,90.0]$. In the region near the region of interest,
more sampling points are placed to increase the amount of data information
and improve fitting accuracy.
The present ChebyKAN model is trained by minimizing the mean squared
error (MSE) between the predicted and reference values
\begin{equation}
\text{MSE}=\frac{1}{N}\sum_{i=1}^{N}\left(E_i^{\text{(pred)}}-
E_i^{\text{(ref)}}\right)^2.
\label{eq:cheykan-nn-1}
\end{equation}
Due to its differentiability and smooth gradient, the MSE is suitable
to serves as loss function during optimization. To assess the global
fitting quality and enable comparison between different models, we
further computed the root mean square error (RMSE). Compared with
conventional multilayer perceptrons (MLP), which form a class of
feedforward artificial NNs with at least three layers, the ChebyKAN
model achieves superior fitting accuracy. A direct comparison of
ChebyKAN with MLPs is performed through the ground-state database,
where the limited-memory Broyden-Fletcher-Goldfarb-Shanno (L-BFGS)
quasi-Newton optimizer and a fixed training duration of $20$ epochs
are employed. In this comparison, the MLP and ChebyKAN are designed with
comparable numbers of parameters. The MLP model adopts a $3$‑$256$‑$256$‑$256$‑$1$
structure with $134401$ parameters, while ChebyKAN uses $K=6$ Chebyshev
polynomials and $3$‑$128$‑$128$‑$1$ NN structure with $118784$ parameters.
Evaluated on the entire ground-state database, the resulting RMSE values
are $1.332\times10^{-3}$ and $2.336\times10^{-4}$ hartree for MLP and
ChebyKAN, respectively, clearly demonstrating the superior performance
of ChebyKAN under a similar parameter budget. This is not surprising.
By integrating these polynomials into the network architecture, the
present ChebyKAN model achieves higher approximation accuracy with
fewer parameters. The orthogonality and recurrence properties of
Chebyshev polynomials ensure that the ChebyKAN remains stable and
efficient.

\subsection{Uncertainty Quantification Automaton\label{sec:uqa-auto}}

The overall procedure of uncertainty quantification automaton contains
four steps. First, a total of $1000$ points are randomly sampled from
the data pool as the initial training set. Second, stepwise hyperparameter
selection within each model family is performed. To accommodate the
small initial dataset, the search begins with relatively small model
configurations. Third, residual-based UQ evaluation is conducted after
the fitting calculation by either ChebyKAN or MLP. Finally, demand-score evaluation is applied
to candidate points in the pool, followed by the selection of new points
for subsequent iterations. The number of sampled points added at each
round is set to $1000$. Letting
the data pool be $\mathcal{D}=\{\mathbi{q}_i,E_i\}_{i=1}^{N}$, we set
$\mathbi{q}_i$ and $E_i$ to be the $i$-th conformation and corresponding
electronic energy, respectively. The automaton first partitions the data
pool $\mathcal{D}$ into a fixed test set $\mathcal{D}_{\mathrm{test}}$,
an initial training set $\mathcal{L}_0$, and an unlabeled candidate pool
$\mathcal{U}_0$, that is $\mathcal{D}=\mathcal{D}_{\mathrm{test}}\cup
\mathcal{L}_0\cup\mathcal{U}_0$. The test set $\mathcal{D}_{\mathrm{test}}$
remains fixed throughout the entire sampling process. Moreover, $\mathcal{L}_t$
denotes the labeled training set at round $t$, while $\mathcal{U}_t$ denotes
the candidate pool that has not yet been selected for training. In each
active learning round, a fixed number of new points $\mathcal{A}_t$ are
selected from $\mathcal{U}_t$, and the sets are updated through
$\mathcal{L}_{t+1}=\mathcal{L}_t\cup\mathcal{A}_t$ and $\mathcal{U}_{t+1}
=\mathcal{U}_t\setminus\mathcal{A}_t$.

In the $t$-th round, for either ChebyKAN or MLP, denoted by $f_{\theta}$,
the automaton first trains a set of candidate architectures on $\mathcal{L}_t$.
The current implementation adopts a Cartesian-grid scan over candidate
hyperparameters. For ChebyKAN, the candidate space is defined as
$\mathcal{H}_{\mathrm{KAN}}=\{(l,w,p):l\in\mathcal{D}_{\mathrm{depth}},
w\in\mathcal{W}_{\mathrm{KAN}},p\in\mathcal{P}_{\mathrm{deg}}\}$, where
$l$ is the hidden-layer depth, $w$ the width of each layer, and $p$ the
Chebyshev polynomial degree. For MLP, the candidate space is
$\mathcal{H}_{\mathrm{MLP}}=\{(l,w):l\in\mathcal{D}_{\mathrm{depth}},
w\in\mathcal{W}_{\mathrm{MLP}}\}$. Each candidate architecture is fully
trained once. The initial hyperparameter grid used in this work is given
in Table \ref{tab:model-select}. The training set is further divided
internally into a fitting subset $\mathcal{T}_t$ and a validation
subset $\mathcal{V}_t$, that is $\mathcal{L}_t=\mathcal{T}_t\cup\mathcal{V}_t$.
The model $f_{\theta}$ is trained on $\mathcal{T}_t$ by minimizing the
MSE given in Equation \eqref{eq:cheykan-nn-1}. During training, the
validation RMSE at each epoch is compared, and after training the model
state with the lowest validation error is restored. This reduces the
influence of terminal overfitting on the comparison.
After all candidate models have been trained in each step, the weighted
validation RMSE is computed for each model and used for selecting the
main model.

Active sampling requires determining which points in the candidate pool
are most valuable for additional data acquisition. This decision is not
based solely on the model-predicted energy error, but rather on the
uncertainty of the model under the support of the current data. The
automaton adopts a structure of main model plus residual UQ. After the
main model is selected in the $t$-th round, the $i$-th residual is defined
on the validation set as $r_i=E_i-f_{\theta_t}(\mathbi{q}_i)$ with
$\{\mathbi{q}_i,E_i\}\in\mathcal{V}_t$. These residuals contain the
systematic errors and local generalization errors of the main model
under the current sampling coverage. The UQ module learns the mapping
$\mathbi{q}\mapsto r(\mathbi{q})$ and provides the predicted residual
mean $\mu_r(\mathbi{q})$ and residual uncertainty $\sigma_r(\mathbi{q})$.
The final corrected prediction is $\tilde{E}(\mathbi{q})=f_{\theta_t}
(\mathbi{q})+\mu_r(\mathbi{q})$, while $\sigma_r(\mathbi{q})$ is used
for candidate-point acquisition scoring and coverage evaluation. We
designed two types of residual UQ backends, namely residual Gaussian
process (GP) and residual expert UQ models. Later in Section \ref{sec:hybrid-uq},
we will turn to details of these model and the present hybrid UQ that integrates them. 

Considering the $t$-th round, the automaton randomly draws a candidate subset
$\mathcal{C}_t$ from the unlabeled pool $\mathcal{U}_t$ to control the
computational cost. For each candidate point $\mathbi{q}\in\mathcal{C}_t$,
the main-model predicted energy $f_{\theta_t}(\mathbi{q})$ and the UQ
standard deviation $\sigma_{\mathrm{UQ}}(\mathbi{q})$ are computed. 
The active sampling score for a candidate point is then defined as
$a(\mathbi{q})=\sigma_{\mathrm{UQ}}(\mathbi{q})w_E(\mathbi{q})$, where
$w_E(\mathbi{q})$ is an energy weight and equal to $1$ or $0.20$ for
the case of $f_{\theta_t}(\mathbi{q})\le E_q$ or $f_{\theta_t}(\mathbf{x})>E_q$,
respectively. Here, $E_q$ is the specified quantile threshold of the
energy distribution in the current training set. The automaton first
selects the subset of candidate points whose $a(\mathbf{x})$ values
lie in the highest quantile region, and then randomly draws a fixed
number of new points from this high-uncertainty subset to be added to
the training set in the next round. This strategy balances exploitation
and exploration. That is, $\sigma_{\mathrm{UQ}}$ promotes exploration
of regions where the model is uncertain, whereas $w_E$ shifts the
sampling focus toward the energy range that is more important for
dynamics calculations. After each training round, to reduce the
influence of errors in physically less relevant regions, the RMSE and
mean absolute error (MAE) are computed in the non-high-energy region
of the fixed test set. The quality of UQ is characterized by the
correlation between uncertainty and error, as well as by coverage.

\subsection{Hybrid Uncertainty Quantification\label{sec:hybrid-uq}}

Having described the uncertainty quantification automaton in Section
\ref{sec:uqa-auto}, now we consider the residual GP and residual experts,
together with the present hybrid UQ that integrates them. Unlike previously
reported active sampling techniques \cite{ele18:17414,chr20:104105,lin20:154104,lin21:2691},
the present hybrid method is implemented through a hybrid criterion that
integrates NN fitting errors, energy percentiles, and the standard
deviation derived from GPR. The present sampling procedure is primarily
driven by the GPR-based uncertainty quantification of the residuals.
Owing to this design, the fitting accuracy demanded of the GPR component
is less stringent than that required in a pure GPR-driven active learning
framework. Consequently, the overall computational cost is substantially
reduced, as the NN serves as the primary fitting engine. In addition,
for either residual GP or residual experts, a single UQ model may
exhibit uncertainty collapse or excessive smoothing. To improve the
stability of active sampling, the automaton with a hybrid mode is
therefore necessary.

The residual GP model
treats the validation residuals as observations of a random function,
where $r(\mathbi{q})\sim\mathrm{GP}(0,K(\mathbi{q},\mathbi{q}'))$, and
then fit the kernel hyperparameters on the validation residual database
$\mathcal{R}_t=\{\mathbi{q}_i,r_i\}_{i\in\mathcal{V}_t}$. In this work,
we adopt a Matern kernel and a white-noise kernel in residual GP
$K(\mathbi{q},\mathbi{q}')=CK_{\mathrm{Matern},\nu=3/2}(\mathbi{q},\mathbi{q}')
+\sigma_r^2\delta_{\mathbi{q},\mathbi{q}'}$. Given a candidate point
$\mathbi{q}_*$, the GP posterior gives the residual mean $\mu_r(\mathbi{q}_*)$
and variance $\sigma_r^2(\mathbi{q}_*)$. In addition, residual expert
can be used as the UQ model. Given the main model (either ChebyKAN or
MLP), a set of small residual models of the same type is further trained
to fit the validation residuals. Letting the $b$-th residual expert be
$g_b(\mathbi{q})\approx r(\mathbi{q})$, the residual mean is taken as
the ensemble average 
\begin{equation}
\mu_r\big(\mathbi{q}\big)=\frac{1}{B}\sum_{b=1}^{B}g_b(\mathbi{q}),
\label{eq:ensemb-ave-gb}
\end{equation}
while the uncertainty is taken as the ensemble standard deviation.
The advantage of residual experts is that they can learn nonstationary,
nonlocal, or model-family-dependent systematic residuals. Their disadvantage
is that they are more computationally expensive than GP, and if the
ensemble members converge to similar solutions, the uncertainty may
be underestimated. Thus, this work also examines whether the active
sampling behavior depends on a specific UQ assumption.

In the hybrid UQ method, the raw standard deviation $\sigma_{\mathrm{raw}}(\mathbi{q})$
given by the UQ backend is computed. A spatial novelty term is then
constructed from the current training points. For a candidate point
$\mathbi{q}$, the distance to its nearest neighbor in the current
training set is computed in the standardized coordinate space, that is
\begin{equation}
d_{\min}\big(\mathbi{q}\big)=
\min_{\mathbi{q}_i\in\mathcal{L}_t}
\big\Vert\hat{\mathbi{q}}-\hat{\mathbi{q}}_i\big\Vert_2.
\label{eq:hybrid-uq-00}
\end{equation}
This distance is normalized by the high quantile $d_{95}$ of the
candidate-point distance distribution,
\begin{equation}
\eta\big(\mathbi{q}\big)=\mathrm{clip}\left(\dfrac{d_{\min}\big(\mathbi{q}\big)}
{d_{95}},0,1\right),
\label{eq:hybrid-uq-01}
\end{equation}
where $d_{95}$ is the $95$-th percentile of the nearest-neighbor distances
from the candidate points to the current training set. In Equation \eqref{eq:hybrid-uq-01},
$\mathrm{clip}(d_{\min}(\mathbi{q})/d_{95},0,1)$ is a clamping function that
restricts $d_{\min}(\mathbi{q})/d_{95}$ to the range $[0,1]$.
The current validation residual RMSE is then used as the energy scale
\begin{equation}
s_r=\left[\dfrac{1}{\vert\mathcal{V}_t\vert}\sum_{i\in\mathcal{V}_t}r_i^2
\right]^{1/2}.
\label{eq:hybrid-uq-02}
\end{equation}
The spatial novelty uncertainty is defined as
\begin{equation}
\sigma_{\mathrm{spatial}}\big(\mathbi{q}\big)=s_r
\alpha_{\mathrm{spatial}}\left(0.15+0.85\eta\big(\mathbi{q}\big)\right),
\label{eq:hybrid-uq-03}
\end{equation}
where $\alpha_{\mathrm{spatial}}(\cdot)$ is a tunable scaling factor,
which is set to $1.0$ by default in this work. In the hybrid mode, if
the coefficient of variation of the raw UQ standard deviation is samller
than a threshold $\epsilon$,
\begin{equation}
\dfrac{\mathrm{std}\big[\sigma_{\mathrm{raw}}\big]}
{\mathrm{mean}\big[\vert\sigma_{\mathrm{raw}}\vert\big]}<\epsilon,
\label{eq:hybrid-uq-04}
\end{equation}
then the UQ is considered to have collapsed. In this case, we directly
adopt the spatial novelty uncertainty, $\sigma_{\mathrm{UQ}}(\mathbi{q})
=\sigma_{\mathrm{spatial}}(\mathbi{x})$. Otherwise, if the raw UQ does
not collapse, we adopt a mixed form
\begin{equation}
\sigma_{\mathrm{UQ}}\big(\mathbi{q}\big)=\left[
\sigma_{\mathrm{raw}}^2\big(\mathbi{q}\big)+\left(
0.35\sigma_{\mathrm{spatial}}\big(\mathbi{q}\big)\right)^2\right]^{1/2}.
\label{eq:hybrid-uq-05}
\end{equation}
This retains the error structure learned by the residual model while
preventing the active sampling process from degenerating into random
selection or simple energy-based filtering when the UQ model fails.

\section{Results and Discussions\label{sec:results}}

\subsection{Automatic Sampling\label{sec:aut-sampl-res}}

\begin{figure}
\centering
\subfigure[\quad Acquisition in automic sampling]{
\includegraphics[width=18cm]{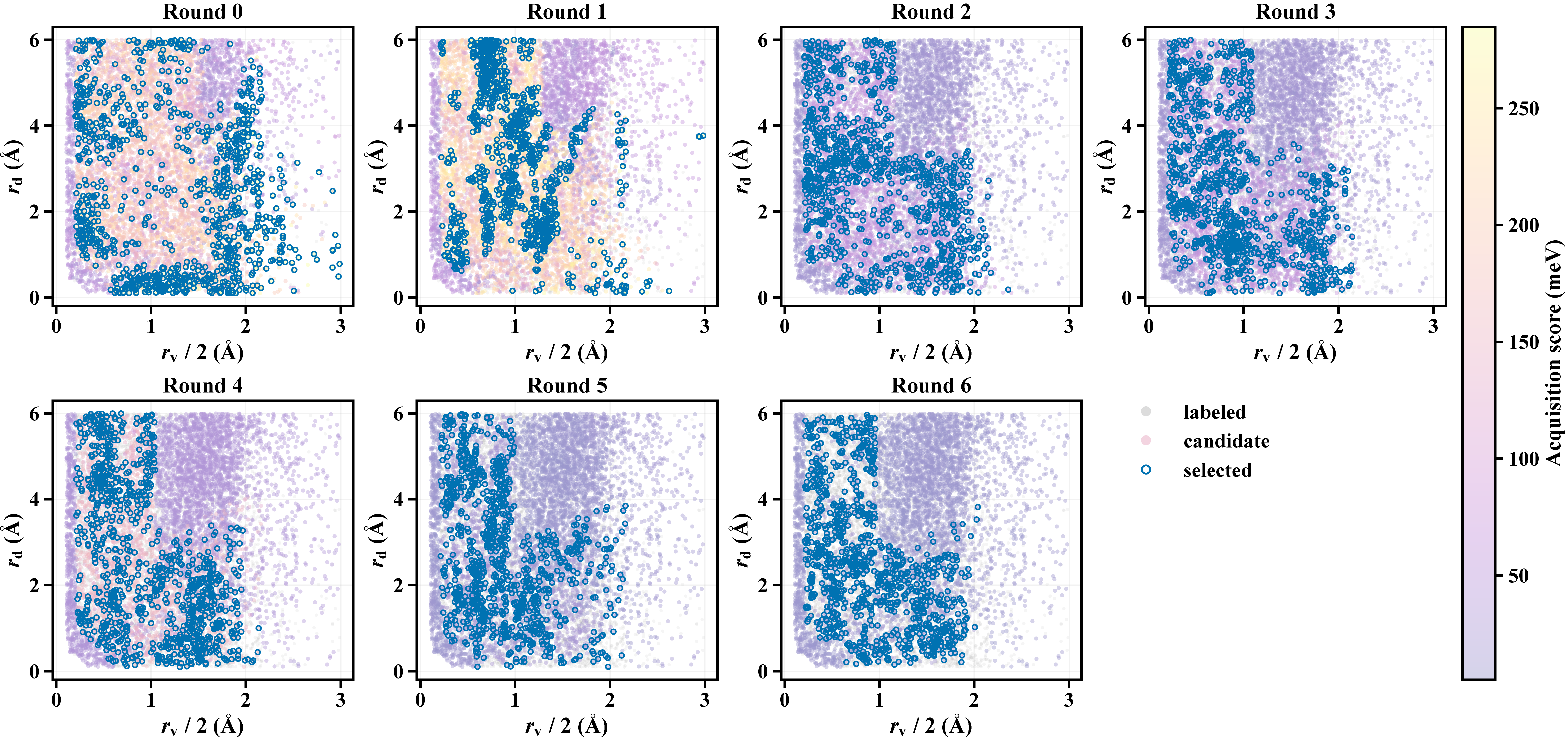}}
\caption{\figfoot}
\label{fig:ana-acti-sampl}
\end{figure}

Before describing the regression results of ChebyKAN, let us first
examine the effectiveness of the uncertainty-driven automatic sampling
workflow. Figure \ref{fig:ana-acti-sampl}(a) summarizes the distribution
of the uncertainty standard deviation for the ground-state element. Each
subplot corresponds to one active learning round. The color of each
point denotes the predicted standard deviation with the color scale
reported in meV. To emphasize the within-round spatial structure of
the uncertainty, each subplot uses an independently normalized color
scale. This representation makes the subtle variations in the low-uncertainty
regime during later sampling stages easier to resolve. As the active
learning procedure proceeds, the high-uncertainty regions contract
overall. The low-energy region associated with the key reaction pathway
becomes progressively better resolved, and its UQ standard deviation
remains at a relatively low level. In contrast, non-critical regions
also show an overall reduction in UQ standard deviation, but consistently
retain larger uncertainty values. These trends indicate that the
automatic sampling procedure continuously supplements underrepresented
regions of the configuration space while preferentially selecting
samples that reduce the uncertainty along reaction-critical regions.
The globally elevated UQ standard deviation observed in the fourth
round is consistent with the fitting fluctuation seen in the training
curve at the same round. This fluctuation may arise from incomplete
optimization within the model family, where a higher overall prediction
error leads to an increased estimated UQ standard deviation. Figure
\ref{fig:ana-acti-sampl}(b) illustrates the capability for spatial
exploration and error-driven sample selection. Instead of selecting
points uniformly from the candidate space, the present sampler
integrates model uncertainty, residual correction, and spatial novelty
information to concentrate new samples in critical regions
that are not yet sufficiently learned by the current model. As the
training set expands round by round, the selected points gradually
shift from a broader coverage of the configuration space toward
localized high-information regions. This indicates that the active
learning workflow can adaptively identify potentially difficult
regions and prioritize sampling in areas that are most relevant to
the construction of an accurate PES. Therefore, Figure \ref{fig:ana-acti-sampl}
demonstrates that the active sampling strategy can improve the
reliability of the PES in critical regions under a limited sampling
budget.

\begin{figure}
\centering
\subfigure[\quad Excited state]{
\includegraphics[width=18cm]{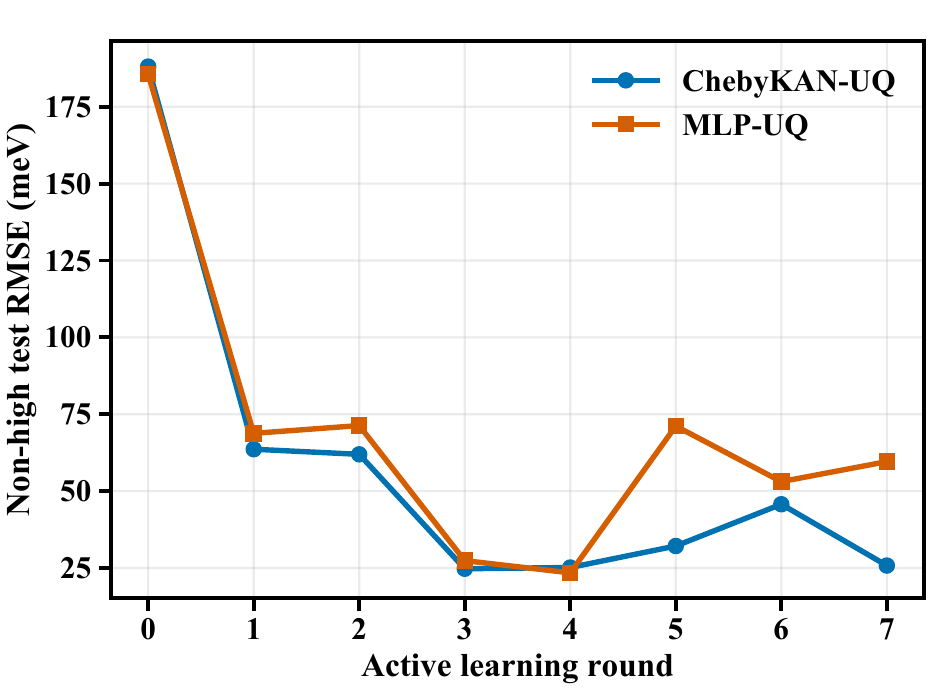}}
\caption{\figfoot}
\label{fig:error-fitting}
\end{figure}

Figure \ref{fig:error-fitting} illustrates the active-learning convergences
of ChebyKAN-UQ and MLP-UQ on the ground-state and excited-state datasets. For
the ground state, the test RMSE of ChebyKAN-UQ decreases rapidly from $284.6$
meV at the zeroth round to $50.7$ meV at the second round, followed by a
continued reduction to $25.6$ meV at the seventh round. One can find that
MLP-UQ also shows a clear decrease in the first few rounds, dropping from
$230.6$ meV to approximately $56\sim70$ meV in later rounds. However, its
late-stage convergence is slower and more fluctuating, ending at $60.7$
meV in the seventh round. Similar situation can be found for the excited
state, where both models achieve a substantial error reduction within the
first three or four rounds. For instance, ChebyKAN-UQ decreases from $188.1$
meV to $24.8$ meV by three rounds and finally remains at $25.8$ meV, while
MLP-UQ decreases from $185.6$ meV to $23.4$ meV by four rounds. Notably,
after the fourth round, both models exhibit a relatively large rebound and
maintain RMSE values higher than those at the fourth round. This may be
attributed to the relatively limited expressive capacity of the initial
model family for the excited-state PES as the number of training points
increases, leading to underfitting behavior. Therefore, the Figure
\ref{fig:error-fitting} demonstrates that the active-learning workflow
can substantially improve the predictive accuracy in the lower-lying
energy region within only a few sampling rounds. This indicates that
the UQ-guided automatic sampling can effectively identify and supplement
configuration regions that are insufficiently learned by the current model.
Moreover, comparing ChebyKAN with MLP-UQ, ChebyKAN-UQ exhibits better
late-stage convergence quality for both electronic states, especially
by achieving lower or more stable final non-high test RMSE values. These
results suggest that combining ChebyKAN-UQ with the active sampling mechanism
can more effectively improve the reliability of the PES model in reaction-critical
energy regions.

\subsection{ChebyKAN Regressions\label{sec:cehby-kan-fit}}

\begin{figure}
\centering
\includegraphics[width=18cm]{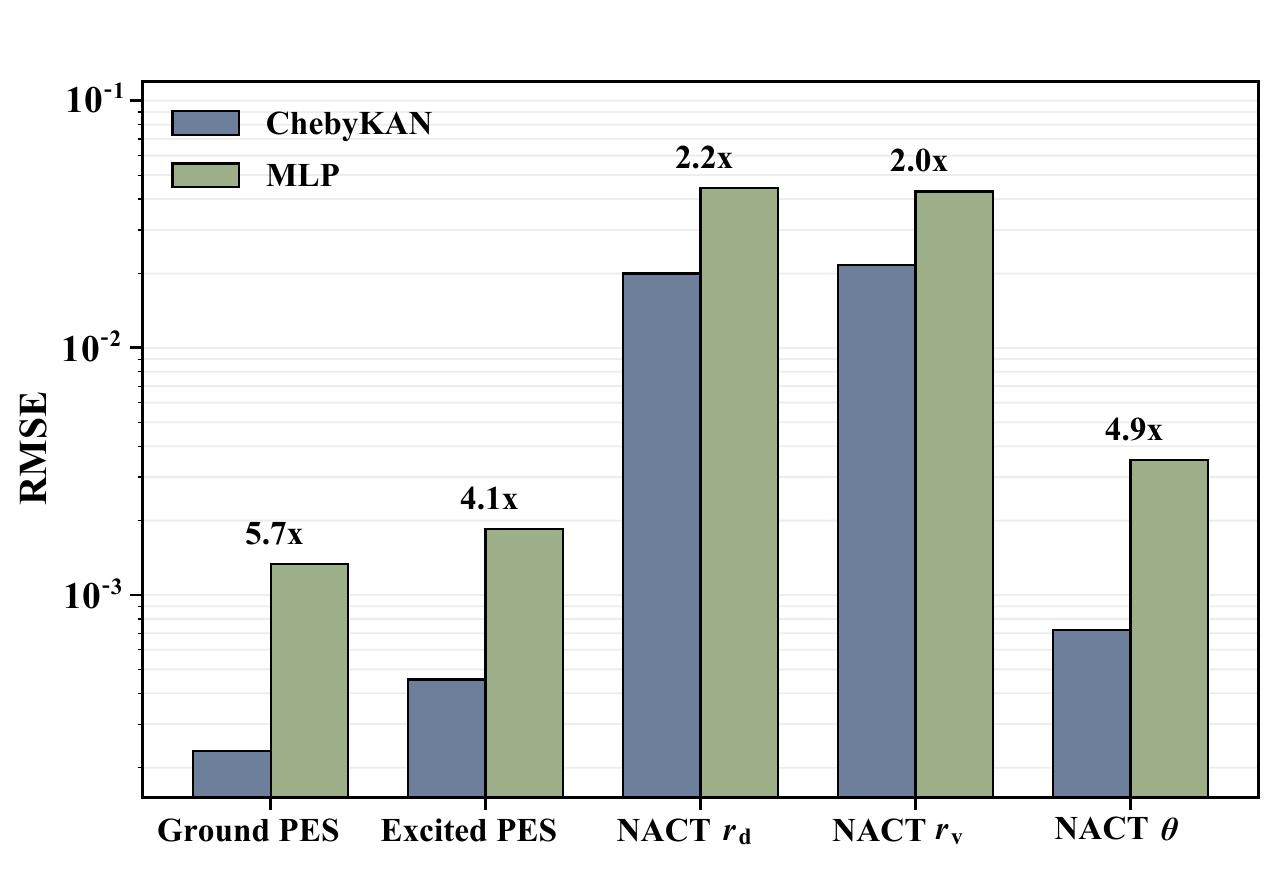}
\caption{\figfoot}
\label{fig:summ-err}
\end{figure}

Based on points sampled in configuration space,
the energy data are computed by the present MRCI/aug-cc-pV5Z calculations.
Using the resulting data sets, the ChebyKAN regressions are then performed
to fit the PES.
We also perform a direct performance comparison between ChebyKAN and
MLP for the ground-state PES using the same optimizer and a fixed
training duration of $20$ epochs. These comprison results are given in
the Supporting Information and indicate that the present ChebyKAN model
achieves superior fitting accuracy. While a larger model can be more
expressive, it does not guarantee superior fitting outcomes. Excessively
complex models, even when they achieve a lower optimization target (say,
the MSE or RMSE value), are prone to overfitting and unnecessarily inflate the
computational cost of subsequent quantum dynamics simulations. To strike
a balance between expressivity and efficiency, we systematically trained
and compared KANs with varying Chebyshev polynomial degrees and hidden
layer widths, as summarized in the Supporting Information. Moreover,
comparison of training results between ChebyKAN and a standard MLP under
similar amount of parameters indicates that ChebyKAN achieves better
fitting accuracy.
For the ground-state and excited-state PESs, as shown in Figure \ref{fig:summ-err},
the MLP errors are approximately $5.7$ and $4.1$ times larger than those 
of ChebyKAN, respectively, under a similar parameter budget. ChebyKAN also 
maintains such advantage for the NACT components. The MLP errors are 
approximately $2.0\sim4.9$ times larger than those of ChebyKAN. These results 
demonstrate that ChebyKAN improves not only the fitting of PES energy values 
but also the learning accuracy of the NACT components.

For both
ground and excited states, raising the polynomial degree from $K=3$ to
$K=6$ substantially improves fitting accuracy. Further widening the hidden
layers from $64$ to $128$ neurons yields a modest enhancement, but expanding
beyond $128$ results in only marginal gains while the parameter count nearly
quadruples, from $(118,784)$ to $(466,944)$. Notably, increasing the
polynomial degree to $K=8$ leads to clear signs of overfitting in the
training results. Based on these results, we select $K=6$ and a
$3$-$128$-$128$-$1$ KAN architecture as the final model for the diagonal
elements, offering a favorable trade-off between accuracy, generalization,
and computational cost. Furthermore, to account for the differing
complexity of the datasets, we inspect two model sizes with polynomial
degrees of $K=3$ and $K=6$, while both retain the same KAN structure of
$3$-$128$-$128$-$1$. By a similar screening procedure, the polynomial
degree of $K=3$ is selected for the coupling elements. This choice
ensures an appropriate balance between fitting accuracy and the risk of
overfitting. The Supporting Information further gives the variation of
the global RMSE for building the ground state PES as increasing the number
of training data. When the number of training data exceeds $12000$, the
RMSE becomes convergence. Within a reliable range, the number of training
data can be optimized to a reasonable value, thereby reducing the
computational cost.

In the Supporting Information (see Figure S4), we illustrate scatter distribution of
the fitting errors (given in the ordinate axis) depending on fitting
targets (given in the abscissa axis). For each element of the potential
matrix, a total of randomly sampled
$\sim10000$ points are employed as training set while the other $2000$
points are used as test set. For constructing the diagonal elements,
the fitting
targets are potential energy values (in hartree), while the fititng
errors are the RMSE values (in meV) on the test set. For fitting the
coupling terms,
the abscissa axis represents the NACT values, while the ordinate axis
represents the fitting errors. According to these error calculations,
the fitting RMSE and MAE values for building the ground state PES are
$3.31$ and $2.38$ meV, respectively, while those for the excited state
PES are $8.09$ and $4.74$ meV. These values indicate the accuracy of
the present ChebyKAN regressions. Moreover, 
the fitting RMSE values for the coupling
terms range from $1.5\times10^{-3}$ au to $3.5\times10^{-2}$ au
leading to the relative RMSE values in the range of $10^{-3}\sim10^{-4}$.
On the other hand, the error distributions displayed in the Supporting Information
are predominantly concentrated
within the range of $\pm2$ meV across the entire energy range, forming a
narrow error band, in particular in constructing the ground state PES.
This observation demonstrates that the present fitting results are both
reliable and accurate. 
The fitting errors are mostly confined to the region below $10$, while in
the region above $10$, where the coupling becomes stronger, the error is
almost zero. This distribution is particularly beneficial for accurately
accounting for the effect of coupling in the subsequent quantum dynamics
calculations (see also Figures S5-S7).

\subsection{Non-Adiabatic Potential\label{sec:matrix-hamiltonian}}

\begin{figure}
    \centering
    \subfigure[\quad 2D contour cut of the excited state PES]{
    \includegraphics[width=18cm]{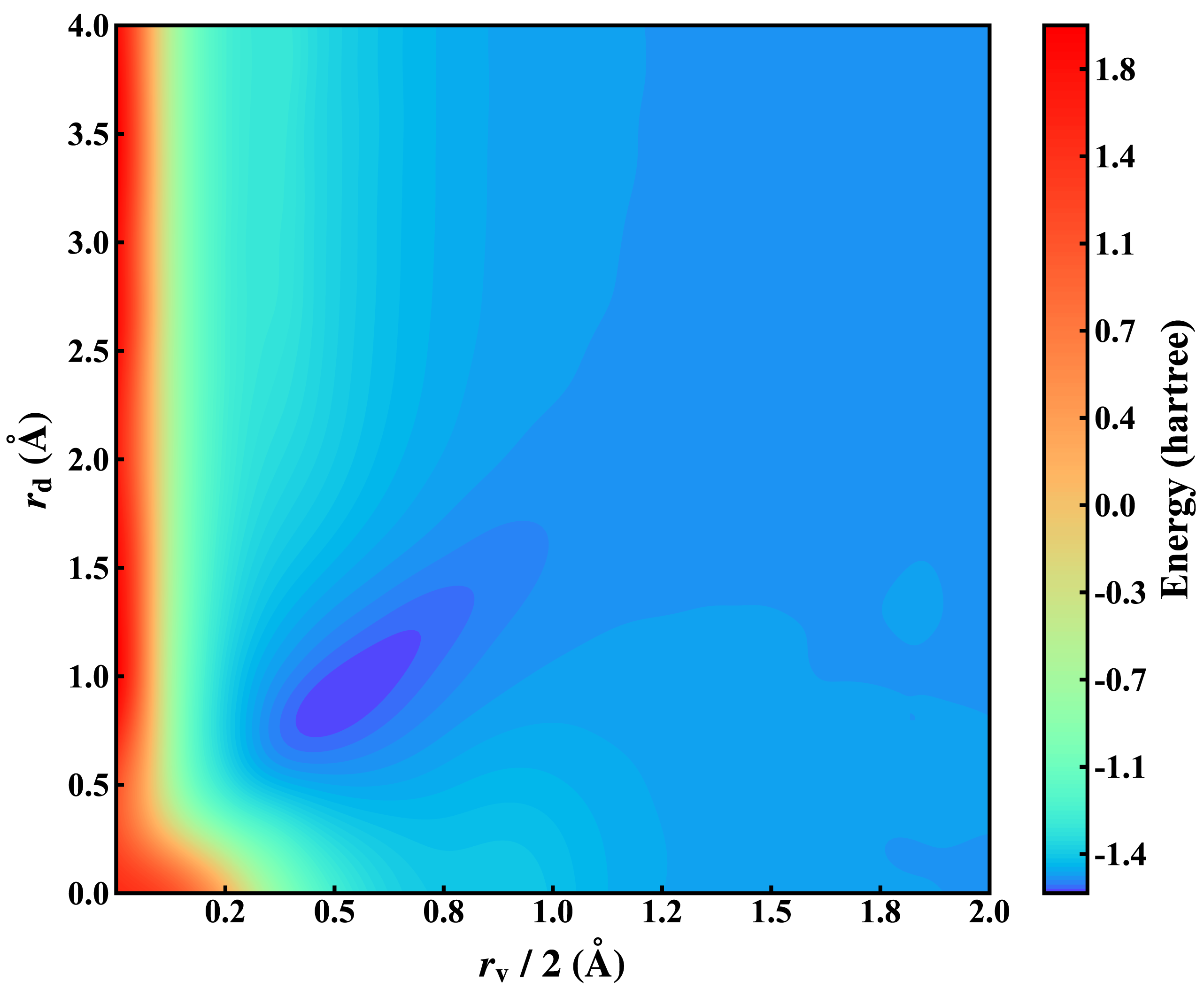}}
    \caption{\figfoot}
    \label{fig:pes-cut}
\end{figure}

Figure \ref{fig:pes-cut} illustrates 2D contour plots of the ground
state PES and the excited state PES with respect to $r_{\mathrm d}$
and $r_{\mathrm v}$ at $\theta=90.0^{\circ}$. In addition to the contour
plots, the colored regions describe the magnitude of the energy. As
explained by the legend on the right, redder colors represent higher
energies, while bluer colors represent lower energies. As given in
Section \ref{sec:cehby-kan-fit}, numerical tests determine that the
ground state and excited state PESs are both fitted through sixth-order
Chebyshev KAN with NN architecture of $3$-$128$-$128$-$1$. Figure
\ref{fig:pes-cut}(a) clearly illustrates reaction characteristic of
the H + H$_2$ system at the ground state. The upper-left corner of
Figure \ref{fig:pes-cut}(a) corresponds to the reactant region,
characterized by a large $r_{\mathrm d}$ value and a small $r_{\mathrm v}$
value that remains near $\sim0.75$ {\AA}. When $r_{\mathrm d}$ is
approaching to $\sim1.5$ {\AA}, the $r_{\mathrm v}$ value progressively
enlarges and reaches $\sim0.95$ {\AA}, indicating that the reaction
occurs. In addition, from the viewpoint of potential energy, Figure
\ref{fig:pes-cut}(a) also clearly reveals the location of the transition
state (TS), which appears as the energy maximum along the reaction
coordinate and corresponds to the saddle point on the PES. Turning to
Figure \ref{fig:pes-cut}(b), one can clearly find that the topology of
the excited-state PES is fundamentally different from that of the ground
state, where no reaction feature is demonstrated. Instead, it is
characterized by a single well-defined energy minimum located in the
region of $r_{\mathrm d}\simeq r_{\mathrm v}\simeq1.0$ {\AA}, indicating
a preference for a more compact, bound configuration without a low-energy
dissociation path.

\begin{figure}
    \centering
    \includegraphics[width=18cm]{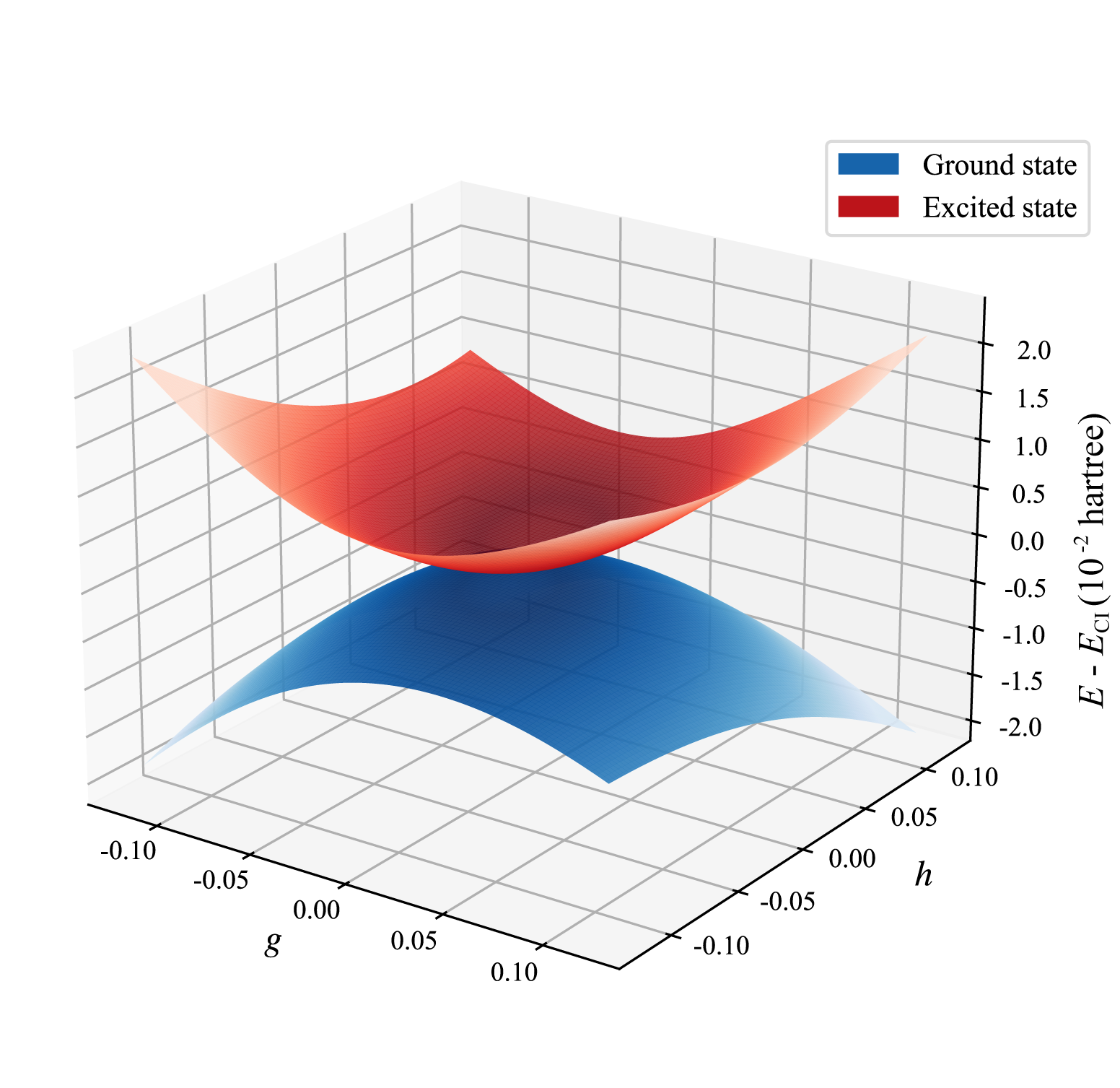}
    \caption{\figfoot}
    \label{fig:CI2_gh_plot}
\end{figure}

In the Supporting Information (see Figure S8), we illustrate 1D contour cus of the
adiabatic potentials of the ground and excited states. 
One can clearly identify two avoided crossing
regions between the contour cuts of these two states, corresponding to
two distinct types, called peaked and sloped types. It is worth noting
that there are peaked and sloped conical intersections (CIs). The former has
the character of a tilted double cone, where the lower (upper) surface
decreases (increases) in all directions from the intersection. The latter
results when both surfaces are monotonically sloped and touch each other
along the slope. 
It should be noted that,
under the $D_{3h}$ symmetry, the H + H$_2$ system exhibits a degeneracy, and the
crossing between the two PESs forms a one-dimensional seam rather than
an isolated point. Under the adiabatic representation, the nature of
the avoided crossing along this seam varies across different regions of
configuration space as the energy changes. Thus, we can find the peaked
and sloped avoided-crossing regions. 
In the Supporting Information (see Figure S9), we illustrate 2D contour cuts of the
energy difference between the ground and excited states around the sloped
and peaked avoided-crossing regions in the $r_{\mathrm d}$-$r_{\mathrm v}$
plane. Figure \ref{fig:CI2_gh_plot} illustrates the 3D contour cut around the peaked avoided-crossing
region in the $g$-$h$ branching space. The positions of avoided crossing
are indicated by asterisks and given in the box in the upper right corner.
The conical avoided crossing can be clearly found in the $g$-$h$ branching
space, where the CI point is located at $\theta=90.0^\circ$ with the
$D_{3h}$ symmetry. 

\begin{figure}
    \centering
    \includegraphics[width=18cm]{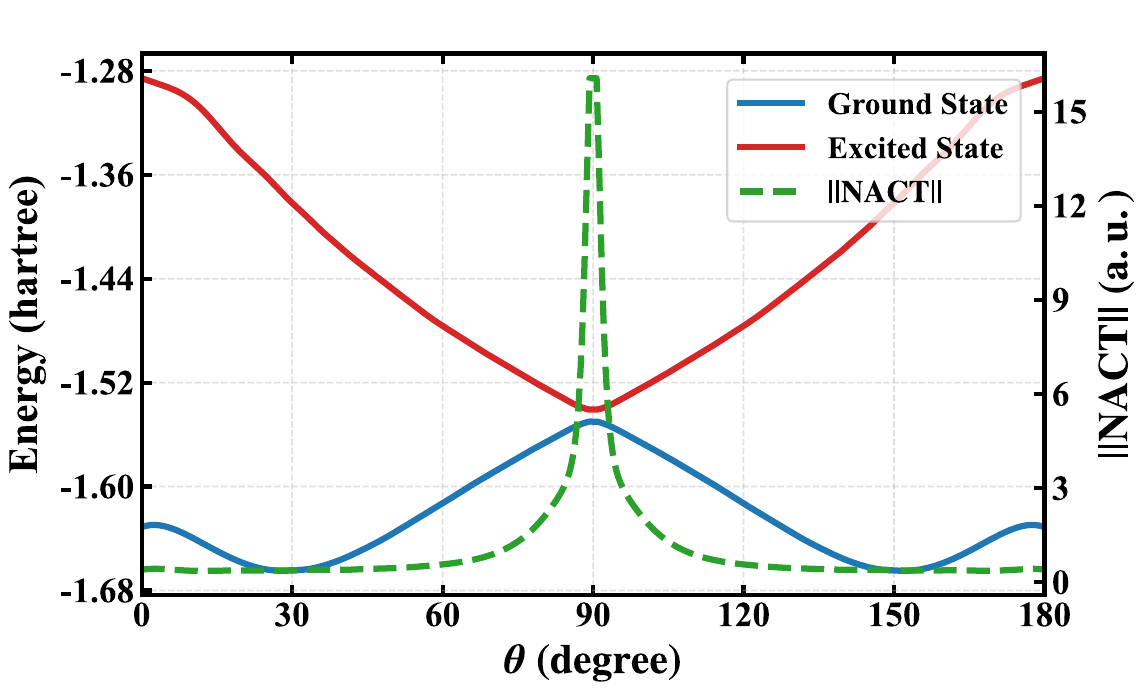}
    \caption{\figfoot}
    \label{fig:energy_NACT_curve}
\end{figure}

The CI of the H + H2 system has been well known by analysis of the
molecular orbitals (MOs). In the $D_{3h}$ symmetry, the three $1s$
atomic orbitals (AOs) of the hydrogen atoms generate a doubly degenerate
set of $E'$ MOs. In the $C_s$ symmetry, the ground state is mainly
characterized by the in-phase combination of the $\sigma$ bonding
orbital of H$_2$ with the $1s$ AO of the incoming hydrogen atom,
whereas the excited state is dominated by the $\sigma^*$ anti-bonding
orbital with the $1s$ AO of the attacking hydrogen atom remaining
non-bonding. Consequently, the excited state is essentially repulsive.
and a $D_{3h}$-symmetric CI connects the two states. This is well
known as the $E\otimes e$ Jahn-Teller effect. It is the $E\otimes e$
Jahn-Teller effect which creates a CI between the two lowest states
along the doubly degenerate $e$-vibrational modes. This is indicates
strong nonadiabatic coupling, which can be clearly found by Figure
\ref{fig:energy_NACT_curve}. In Figure \ref{fig:energy_NACT_curve},
we show energy curves of the ground state (blue solid line) and excited state
(red solid line) as functions of $\theta$, together with the nonadiabatic
coupling (green dashed line) as function of $\theta$. The $r_{\mathrm d}$
and $r_{\mathrm v}$ values are fixed at $r_{\mathrm v}=1.31$ {\AA} and
$r_{\mathrm d}=0.76$ {\AA}. 
The value of the coupling term exhibits a significant abrupt change at
the CI point, that is $\theta=90.0^{\circ}$.

\subsection{Reaction Dynamics\label{sec:wave-prop}}

Next, let us turn to numerical results of the present quantum dynamics
calculations. Given in Table \ref{tab:num-prop-detail} are numerical
details. For the present benchmark, H + H$_2$, the center-of-mass (COM)
of the H$_2$ molecule is G, while the coordinates are bond length
$r_{\mathrm v}$, dissociation length $r_{\mathrm d}$, and their included
angle $\theta$. Since G is located at the center of $r_{\mathrm v}$ making
$r_{\mathrm v}/2$ a convenient parameter of conformation.
The ranges of the coordinates are
set to be $0.529\:{\mathrm\AA}\leq r_{\mathrm d}\leq 4.782\:{\mathrm\AA}$,
$0.317\:{\mathrm\AA}\leq r_{\mathrm v}\leq 3.300\:{\mathrm\AA}$, and
$0^{\circ}\leq\theta\leq180^{\circ}$, while $108$, $68$, and $45$ grid
points are used for $r_{\mathrm d}$, $r_{\mathrm v}$, and $\theta$,
respectively. Moreover, as given in Table \ref{tab:num-prop-detail},
to ensure convergence in both MCTDH and POTFIT calculations, the
single-set of SPF sizes $r_{\mathrm d}/r_{\mathrm v}/\theta=60/60/40$ 
and SPP size of $r_{\mathrm d}/r_{\mathrm v}/\theta=c/50/30$ ($c$
means contracted mode) are used, respectively. Using such SPP size,
there are $1500$ terms in the {\it potfitted} function. Moreover, we
further define the relevant grid points \cite{bec00:1} as those with
potential energy being less than $-1.0$ eV for the ground state PES,
while the relevant grid points are defined as those with $r_{\mathrm d}<1.5$ {\AA}
for other elements. On these relevant grid points, the RMSEs of POTFIT
are smaller than $0.5$ meV, which is much smaller than the PES fitting
RMSEs (smaller than $5$ meV). Even on all grid points, the RMSEs of
POTFIT are around $1$ meV that is still much smaller than the fitting
RMSEs. This clearly indicates the convergence of the present POTFIT
calculations.

Having constructed the whole Hamiltonian operator and set the CAP, the initial wave functions
$\vert\Psi(0)\rangle$ are constructed by MCTDH relaxation calculations
based on the guess function (see also Table \ref{tab:num-prop-detail}).
The vibration ground state of the H$_2$ molecule is used as reactants,
while the third H atom is located at $r^{(0)}_{\mathrm d}=4.5$ {\AA} 
with an initial momentum of $p=8.0$ au moving towards H$_2$, as given by
Table \ref{tab:num-prop-detail}. Shown in Supporting Information are
reduced densities of time-dependent wave functions on the
$r_{\mathrm d}$-$r_{\mathrm v}$ plane with $\theta=90.0^{\circ}$.
Here we would like to emphasize that in computing the initial wave 
functions an artificial harmonic potential $0.01\times(r_{\mathrm d}-r^{(0)}_{\mathrm d})^2$
hartree is used to fix the third H atom at $r^{(0)}_{\mathrm d}=4.5$
{\AA}. Finally, the relaxed wave function is operated by the operator
$\exp(\mathrm{i}pr_{\mathrm d})$ to assign a momentum of $-8.0$ au,
where the minus sign means that H moves towards H$_2$. With the initial
wavepacket, the MCTDH propagation calculations are performed to compute
the time-dependent wavepacket. The reaction dynamics is then obtained
through analysis of the time-dependent wavepacket adsorbed by the CAP.

\begin{figure}
    \centering
    \includegraphics[width=18cm]{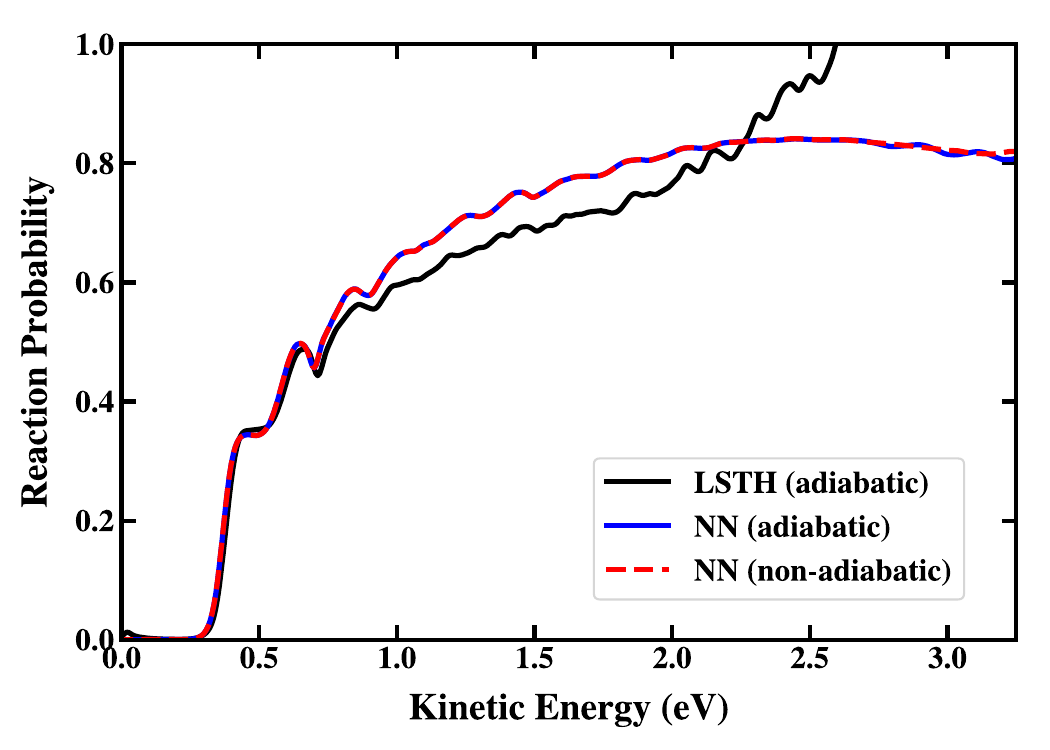}
    \caption{\figfoot}
    \label{fig:reaction_prob}
\end{figure}

The Supporting Information (see Figure S10) illustrates time-dependent electronic-state
populations computed by the initial wave function at either the ground
state or the excited state. The abscissa and ordinate represent time
(in fs) and populations. The blue and red lines represent populations
of the ground and excited states, respectively, corresponding to the
left (blue words)and right (red words) axes. 
The ground state hardly transfers to the excited state,
with the ground-state population remaining close to $100\%$ even after
$200$ fs. This is attributed not only to energy constraints but also to
the weak coupling between the two states, as evidenced by those from
the initial wave function at the excited state. 
The excited state still retains approximately $85\%$ of its
population after $500$ fs, confirming that the coupling plays only a
minor role in the dynamics of this system.
Given in Figure\ref{fig:reaction_prob} are the present reactive probability for
the H + H$_2$ reaction. In Figure \ref{fig:reaction_prob} we set the potential 
energy of reactant H + H$_2$ (denoted by $U_R$) as the original point 
of the kinetic-energy axis, that is set $U_R=0$. On the basis of each 
of the present ChebyKAN PESs as well as the previously reported LSTH PES,
the reactive probabilities are computed by the MCTDH method. In Supporting
Information we give a movie for the motions of the wave packet on the PES
to show the reaction process as well as the time scale of this reaction.
From Figure \ref{fig:reaction_prob}, one can also find that at lower-lying
kinetics energy of smaller than $1.0$ eV all of the dynamics results are
close to each other which implies the accuracy of the present PESs. At
higher-lying kinetics energy region of larger than $1.0$ eV, calculations
on these PESs predict reasonable reactive probabilities, except for rather
larger probability based on the present PES. Moreover, the reactive
probabilities computed by the adiabatic and non-adiabatic PESs are close
to each other. This also implies that the coupling in the H + H$_2$ system
hardly influence the reaction dynamics.

\section{Conclusions\label{sec:con}}

In this work, we propose and implement the ChebyKAN model and develop
the uncertainty-quantification-driven active sampling technique to
sample the conformation data in the configuration space, where a hybrid
criterion is employed integrating NN fitting errors, energy percentiles,
and the standard deviation derived from GPR.
As a benchmark, the non-adiabatic PES of the H + H$_2$ system is
constructed. For each element of the potential matrix, a total of $\gtrsim 10^4$ data
are computed at the MRCI/aug-cc-pV5Z level in conjunction with rather
large active space. Utilizing the data sets computed in this work, the ChebyKAN method is systematically
tailored to the PES construction through a thorough optimization of its hyperparameters.
Its numerical performance is then compared against that of established NN
techniques. A systematic validation of the accuracy of ChebyKAN is performed.
Our numerical results precisely located the conical intersection and
faithfully reproduced the characteristic features of the non-adiabatic
dynamics of H + H$_2$, which are fully consistent with those reported
in earlier studies. Based on the present PES, the time-dependent nuclear
Schr{\"o}dinger equation is solved by the MCTDH method, where all of 
elements of the potential matrix have been re-fitted in the SOP form
by the POTFIT calculations. By the wavepacket propagation, one can
find that the coupling terms hardly influence the quantum dynamics of
the ground state. Then, a deep discussion of the non-adiabatic quantum
dynamics is given.

\section*{Supplementary Material}

See Supporting Information at http://dx.doi.org/XXX for
numerical comparison of the present ChebyKAN model with MLP
and convergence inspection of the hyperparameters in the present
ChebyKAN model.

\section*{Interest Statement}

The authors declare that they have no known competing financial interests
or personal relationships that could have appeared to influence the work
reported in this paper.

\section*{Data Availability}

All data have been reported in this work. The data supporting this article
have been included as part of the Supplementary Information.

\section*{Acknowledgements}

The financial supports of National Natural Science Foundation of China
(Grant No. 22273074) and Fundamental Research Funds for the Central
Universities (Grant Nos. 2025JGZY34 and 2025KCW017) are gratefully
acknowledged. We are grateful to the anonymous reviewers for their
thoughtful suggestions.


\clearpage
\begin{table}
\centering
\caption{Candidate hyperparameter grids used for model selection in
each active-learning round. The first column gives models in the
present work. The second column gives the hyperparameter sets. The
third column gives the corresponding values.}
\begin{tabular}{lclr}
\hline
Model &~~& Hyperparameter set & Candidate values \\
\hline
ChebyKAN
&& $\mathcal{D}_{\mathrm{depth}}$ & $\{2,3\}$ \\
&& $\mathcal{W}_{\mathrm{KAN}}$   & $\{20,32,64,128\}$ \\
&& $\mathcal{P}_{\mathrm{deg}}$   & $\{2,3,4,6,8\}$ \\
MLP
&& $\mathcal{D}_{\mathrm{depth}}$ & $\{2,3\}$ \\
&& $\mathcal{W}_{\mathrm{MLP}}$ & $\{32,64,96,128,192,256\}$ \\
\hline
\end{tabular}
\label{tab:model-select}
\end{table}

\clearpage
\begin{sidewaystable}
\caption{The time-independent primitive basis functions, number and range
of grid points, and sizes of SPFs and SPPs used in both POTFIT and MCTDH
calculations, as well as parameters of the initial wave function for the
propagation. The definitions of the coordinates (indicated in the first
column) are given in Figure S1 of the Supporting Information. The second column describes
the primitive basis functions, which underlay the DVR. The third column
gives the number of the grid points. The fourth column gives sizes of
SPFs and SPPs (in parenthesis) for MCTDH and POTFIT calculations,
respectively, where $c$ means contracted mode while the single-set
fashion is adopted in the present two-state MCTDH calculations. The
fifth column gives the range of the grids in {\AA}ngstrom or
degree. The sixth column gives the symbol of the one-dimensional (1D)
function for each coordinate in the initial wave function. The other
columns give the parameters for these 1D functions, including width
of the GAUSS function and the initial quantum numbers ($j_{\mathrm{ini}}$
and $m_{\mathrm{ini}}$) of the angular functions.}
\begin{center}
 \begin{tabular}{llllllllrlrlrlrlr}
  \hline
 Coordinates &~~& \multicolumn{7}{c}{Basis function}  
 &~~~~~~& \multicolumn{7}{c}{Initial wave function}  \\ \cline{3-9} \cline{11-17}
 &~~& Symbol
 \footnote{Symbols ``SIN'' and ``LEG'' stand for the sine DVR and Legendre DVR \cite{suk01:2604}
for angular coordinate, respectively.}
 &~~& Grid points &~~& Size &~~& Range of the grids &~~~~~~& Symbol
 \footnote{Symbols ``GAUSS'' designates the choice of Gaussian function as initial SPFs. 
Symbol ``EIGENF'' means eigenfunction of a specified potential which, in this work,
is the potential energy of the H$_2$ molecule. Symbol ``LEG'' denotes Legendre 
function to specify the initial wave function.}
 &~~& Position &~~& Momentum &~~& Parameters  \\
\hline
$r_{\mathrm{d}}$&&SIN&&108&&60 ($c$)&&$[0.529,4.782]$&&GAUSS&&$4.5$&&$-8.0$&&
$W_{\mathrm{GAUSS}}=0.25$   \\
$r_{\mathrm{v}}$&&SIN&&68&&60 (50)&&$[0.317,3.300]$&&EIGENF&& --  && --  && ground state \\
$\theta$&&LEG&&45&&40 (30)&&$[0.000,180.000]$&&LEG&& --- && --- &&
$j_{\mathrm{ini}}=j$, all $j$ values are used  \\
\hline
 \end{tabular}
  \end{center}
   \label{tab:num-prop-detail}
    \end{sidewaystable}
            
\clearpage
\section*{Figure Caption}

\figcaption{fig:ana-acti-sampl}{
Analysis of the automatic sampling based on (a) uncertainty quantification
standard deviation and (b) acquisition in sampling. Subfigure (a)
summarizes the spatial distribution of the uncertainty standard deviation
which is predicted by the ChebyKAN model on a fixed test set during active
learning rounds for the ground-state PES. Each subplot corresponds to an
active learning round. The color of each point denotes the predicted standard
deviation, with the color scale given in meV. Subfigure (b) illustrates
the distributions of candidate samples (light-colored points), previously
labeled training samples (gray points), and newly recommended samples
(blue points) during active learning rounds. The color of light-colored
points indicates the acquisition score.
}

\figcaption{fig:error-fitting}{
Overall fitting accuracies of ChebyKAN and MLP during active sampling
for (a) the ground-state PES and (b) the excited-state PES. The horizontal
axis denotes the active learning round. The vertical axis reports the
non-high-energy test RMSE (in meV). The ChebyKAN model adopts
a 3-128-128-1 architecture, with the highest Chebyshev polynomial degree
set to $K=6$. The MLP uses a 3-256-256-256-1 architecture with a comparable
parameter budget.
}

\figcaption{fig:summ-err}{
Comparison the overall fitting accuracy of ChebyKAN with MLP, including
two adiabatic PESs and NACT components. Each element of the potential
matrix is represented by a pair of bars, where the blue-gray bars denote
the ChebyKAN results and the green bars denote the MLP results. The
vertical axis reports the RMSE on a logarithmic scale. As given in the
main text, ChebyKAN adopts a structure of 3-128-128-1 with the highest
Chebyshev polynomial degree of $K=6$ for the two adiabatic PESs and $K=3$
for the coupling terms. The MLP uses a NN structure of 3-256-256-256-1
with a comparable parameter budget.
}

\figcaption{fig:pes-cut}{
Two-dimensional (2D) contour plots of the (a) ground state and (b) excited state
PESs with respect to $r_{\mathrm d}$ and $r_{\mathrm v}$ at $\theta=90.0^{\circ}$.
Redder colors represent higher energies, while bluer colors represent lower
energies. The ground state and excited state PESs are both fitted through
sixth-order Chebyshev KAN with the NN architecture of $3-128-128-1$.
}

\figcaption{fig:CI2_gh_plot}{
Three-dimensional (3D) contour cut of the
potential energy around the peaked CI point in the $g-h$ branching space. 
The conical avoided crossing of the peaked CI point can be clearly found in the $g$-$h$ branching
space, where the peaked CI point is located at $r_{\mathrm d}=1.31$ {\AA},
$r_{\mathrm v}=0.76$ {\AA}, and $\theta=90.0^\circ$. The ordinate represents
potential energy, with zero set at the CI point. The blue and red surfaces represent the
ground and excited state, respectively. 
}

\figcaption{fig:energy_NACT_curve}{
Energy curves of the ground state (blue solid line) and excited state
(red solid line) as functions of $\theta$, together with the nonadiabatic
coupling term (green dashed line) as function of $\theta$, where the
$r_{\mathrm d}$ and $r_{\mathrm v}$ values are fixed at those for the
peaked CI point, namely $r_{\mathrm v}=1.31$ {\AA} and $r_{\mathrm d}=0.76$
{\AA}. The value of the coupling term exhibits a significant abrupt
change at the CI point, that is $\theta=90.0^{\circ}$.
}

\figcaption{fig:reaction_prob}{
Energy-dependent reaction probabilities versus kinetic energy (in eV)
for the H$'$ + H$_2$ $\to$ HH$'$ + H reaction where 3D two-state
Hamiltonian operator is used to perform the 3D MCTDH calculations.
The black solid line represents reaction probability computed through
the LSTH adiabatic PES. The blue and red lines represent reaction
probabilities computed through the present NN adiabatic and non-adiabatic
PESs, respectively.
}

\clearpage
\begin{figure}
\centering
\subfigure[\quad Uncertainty quantification in automic sampling]{
\includegraphics[width=18cm]{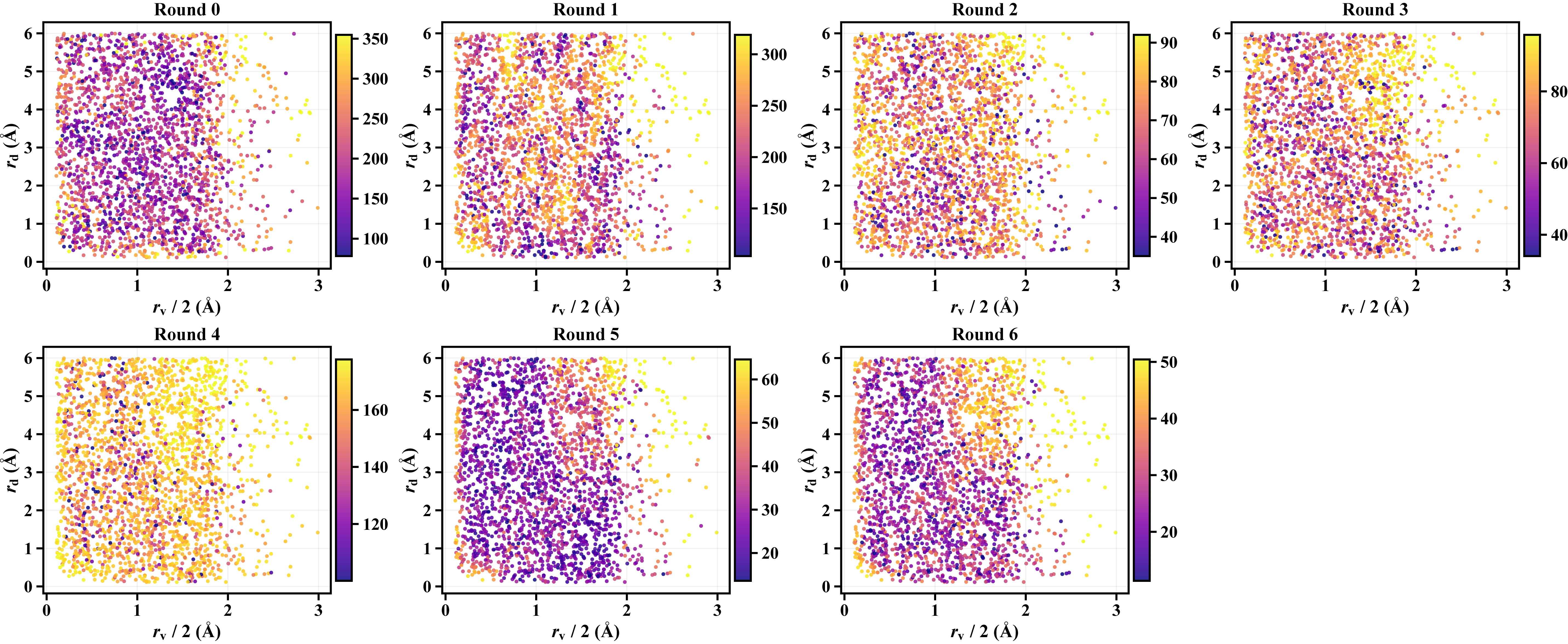}}
\end{figure}

\clearpage
\begin{figure}
\centering
\subfigure[\quad Ground state]{
\includegraphics[width=18cm]{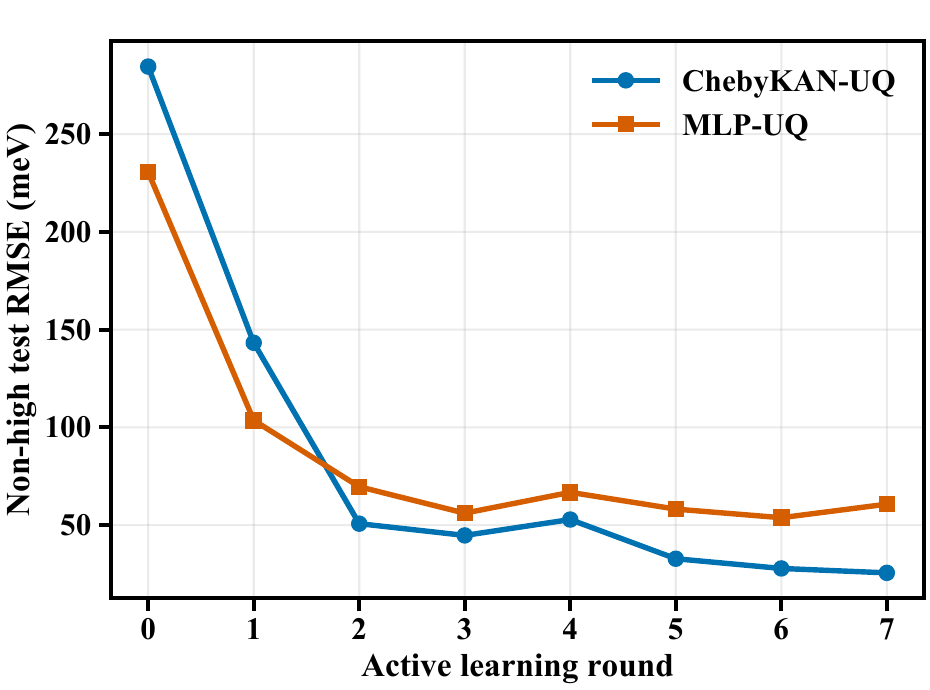}}
\end{figure}

\clearpage

\clearpage
\begin{figure}
    \centering
    \subfigure[\quad 2D contour cut of the ground state PES]{
    \includegraphics[width=18cm]{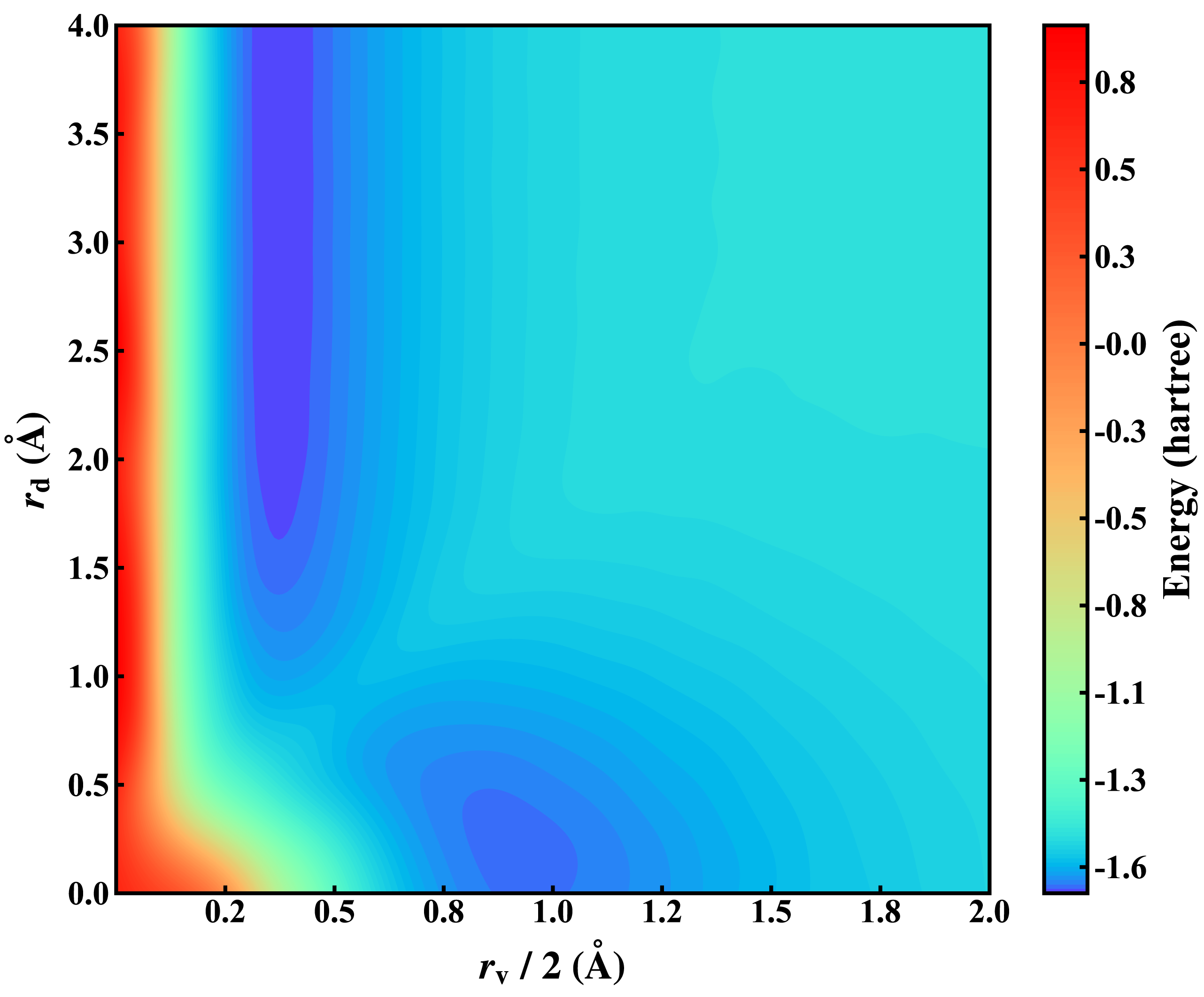}}
    \end{figure}
\clearpage

\clearpage

\clearpage

\clearpage

\clearpage

\begin{thebibliography}{10}
\ifnum\language=1 \def\biband{und}\else\def\biband{and}\fi

\bibitem{eve05:134110}
{\rm C.~R. Evenhuis, X.~Lin, D.~H. Zhang, D.~R. Yarkony,  \biband~ M.~A.
  Collins}.
\newblock Interpolation of diabatic potential-energy surfaces: Quantum dynamics
  on {\it ab initio} surfaces.
\newblock {\em J.~Chem.\ Phys. \bf 123\/} (2005), 134110.

\bibitem{beh07:014705}
{\rm J.~Behler, S.~Lorenz,  \biband~ K.~Reuter}.
\newblock Representing molecule-surface interactions with symmetry-adapted
  neural networks.
\newblock {\em J.~Chem.\ Phys. \bf 127\/} (2007), 014705.

\bibitem{beh07:146401}
{\rm J.~Behler \biband~ M.~Parrinello}.
\newblock Generalized neural-network representation of high-dimensional
  potential-energy surfaces.
\newblock {\em Phys.\ Rev.\ Lett. \bf 98\/} (2007), 146401.

\bibitem{beh11:17930}
{\rm J.~Behler}.
\newblock Neural network potential-energy surfaces in chemistry: a tool for
  large-scale simulations.
\newblock {\em Phys.\ Chem.\ Chem.\ Phys. \bf 13\/} (2011), 17930--17955.

\bibitem{beh15:1032}
{\rm J.~Behler}.
\newblock Constructing high-dimensional neural network potentials: A tutorial
  review.
\newblock {\em Int.\ J.\ Quant.\ Chem. \bf 115\/} (2015), 1032--1050.

\bibitem{che15:1241}
{\rm J.~Chen \biband~ D.~H. Zhang}.
\newblock Construction of molecular reactive potential energy surfaces based on
  neural networks.
\newblock {\em Sci. Sin. Chim. \bf 45\/} (2015), 1241--1253.

\bibitem{bar18:041048}
{\rm A.~P. Bart{\'o}k, J.~Kermode, N.~Bernstein,  \biband~ G.~Cs{\'a}nyi}.
\newblock Machine learning a general-purpose interatomic potential for silicon.
\newblock {\em Phys.\ Rev.~X \bf 8\/} (2018), 041048.

\bibitem{kam18:241702}
{\rm A.~Kamath, R.~A. Vargas-Hern{\'a}ndez, R.~V. Krems, T.~Carrington,
  \biband~ S.~Manzhos}.
\newblock Neural networks vs gaussian process regression for representing
  potential energy surfaces: A comparative study of fit quality and vibrational
  spectrum accuracy.
\newblock {\em J.~Chem.\ Phys. \bf 148\/} (2018), 241702.

\bibitem{wan19:4513}
{\rm H.~Wang \biband~ R.~P.~A. Bettens}.
\newblock Modelling potential energy surfaces for small clusters using shepard
  interpolation with gaussian-form nodal functions.
\newblock {\em Phys.\ Chem.\ Chem.\ Phys. \bf 21\/} (2019), 4513--4522.

\bibitem{fu19:145}
{\rm B.~Fu, J.~Chen, T.~Liu, K.~Shao,  \biband~ D.~H. Zhang}.
\newblock {Highly Accurately Fitted Potential Energy Surfaces for Polyatomic
  Reactive Systems}.
\newblock {\em Acta\ Phys.-Chim.\ Sin. \bf 35\/} (2019), 145--157.

\bibitem{gua18:823}
{\rm Y.~Guan, S.~Yang,  \biband~ D.~H. Zhang}.
\newblock Construction of reactive potential energy surfaces with gaussian
  process regression: active data selection.
\newblock {\em Mol.\ Phys. \bf 116\/} (2018), 823--834.

\bibitem{son19:114302}
{\rm Q.~Song, Q.~Zhang,  \biband~ Q.~Meng}.
\newblock Neural-network potential energy surface with small database and high
  precision: A benchmark of the h + h2 system.
\newblock {\em J.~Chem.\ Phys. \bf 151\/} (2019), 114302.

\bibitem{son22:1983}
{\rm Q.~Song, X.~Zhang, Z.~Miao, Q.~Zhang,  \biband~ Q.~Meng}.
\newblock Unified regression models in fitting potential energy surfaces for
  quantum dynamics.
\newblock {\em J.~Math.\ Chem. \bf 60\/} (2022), 1983--2012.

\bibitem{son22:11128}
{\rm Q.~Song, X.~Zhang, D.~Pel{\'a}ez,  \biband~ Q.~Meng}.
\newblock Direct canonical-polyadic-decomposition of the potential energy
  surface from discrete data by decoupled gaussian process regression.
\newblock {\em J.~Phys.\ Chem.\ Lett. \bf 13\/} (2022), 11128--11135.

\bibitem{son24:597}
{\rm Q.~Song, X.~Zhang, Z.~Miao,  \biband~ Q.~Meng}.
\newblock Construction of mode-combination hamiltonian under the gridbased
  representation for the quantum dynamics of oh + ho$_2$ $\to$ o$_2$ + h$_2$o.
\newblock {\em J.~Chem.\ Theory Comput. \bf 20\/} (2024), 597--613.

\bibitem{att97:1069}
{\rm J.~G. Attali \biband~ G.~Pag{\'e}s}.
\newblock {Approximations of functions by a multilayer perceptron: a new
  approach}.
\newblock {\em Neural Networks \bf 10\/} (1997), 1069--1081.

\bibitem{sch07:253}
{\rm A.~M. Sch{\"a}fer \biband~ H.-G. Zimmermann}.
\newblock Recurrent neural networks are universal approximatiors.
\newblock {\em Int. J. Neu. Sys. \bf 17\/} (2007), 253--263.

\bibitem{hed19:123}
{\rm D.~He, T.~Zhang, J.~Yuan,  \biband~ M.~Wang}.
\newblock {A new potential energy surface of the LiHO$^+$ system and the
  dynamics studies of the O + LiH$^+$ reaction}.
\newblock {\em Chem.\ Phys.\ Lett. \bf 715\/} (2019), 123--128.

\bibitem{xu19:1151}
{\rm X.~Xu, J.~Chen, S.~Liu,  \biband~ D.~H. Zhang}.
\newblock An {\it ab initio} based global potential energy surface for the
  sh$_3$ system and full-dimensional state-to-state quantum dynamics study for
  the {H$_2$ + HS $\to$ H$_2$S + H} reaction.
\newblock {\em J.~Comp.\ Chem. \bf 40\/} (2019), 1151--1160.

\bibitem{lux18:23095}
{\rm X.~Lu, K.~Shao, B.~Fu, X.~Wang,  \biband~ D.~H. Zhang}.
\newblock An accurate full-dimensional potential energy surface and
  quasiclassical trajectory dynamics of the {H + H$_2$O$_2$} two-channel
  reaction.
\newblock {\em Phys.\ Chem.\ Chem.\ Phys. \bf 20\/} (2018), 23095--23105.

\bibitem{wan18:17960}
{\rm S.~Wang, Z.~Yang, J.~Yuan,  \biband~ M.~Chen}.
\newblock New diabatic potential energy surfaces of the {NaH${_2}$} system and
  dynamics studies for the {Na(${}^3P$) + H$_2$ $\to$ NaH + H} reaction.
\newblock {\em Sci.\ Rep. \bf 8\/} (2018), 17960.

\bibitem{gua17:224307}
{\rm Y.~Guan, B.~Fu,  \biband~ D.~H. Zhang}.
\newblock Construction of diabatic energy surfaces for {LiFH} with artificial
  neural networks.
\newblock {\em J.~Chem.\ Phys. \bf 147\/} (2017), 224307.

\bibitem{zha16:135}
{\rm D.~H. Zhang \biband~ H.~Guo}.
\newblock Recent advances in quantum dynamics of bimolecular reactions.
\newblock {\em Ann.\ Rev.\ Phys.\ Chem. \bf 67\/} (2016), 135--158.

\bibitem{li13:204103}
{\rm J.~Li, B.~Jiang,  \biband~ H.~Guo}.
\newblock Permutation invariant polynomial neural network approach to fitting
  potential energy surfaces. {II}. {F}our-atom systems.
\newblock {\em J.~Chem.\ Phys. \bf 139\/} (2013), 204103.

\bibitem{jia13:054112}
{\rm B.~Jiang \biband~ H.~Guo}.
\newblock Permutation invariant polynomial neural network approach to fitting
  potential energy surfaces.
\newblock {\em J.~Chem.\ Phys. \bf 139\/} (2013), 054112.

\bibitem{jia14:034109}
{\rm B.~Jiang \biband~ H.~Guo}.
\newblock Permutation invariant polynomial neural network approach to fitting
  potential energy surfaces. {III. M}olecule-surface interactions.
\newblock {\em J.~Chem.\ Phys. \bf 141\/} (2014), 034109.

\bibitem{bow11:8094}
{\rm J.~M. Bowman, G.~Czak{\'o},  \biband~ B.~Fu}.
\newblock High-dimensional ab intio potential energy surafces for reaction
  dynamics calculations.
\newblock {\em Phys.\ Chem.\ Chem.\ Phys. \bf 13\/} (2011), 8094--8111.

\bibitem{cza12:044307}
{\rm G.~Czak{\'o} \biband~ J.~M. Bowman}.
\newblock Accurate ab initio potential energy surface, thermochemistry, and
  dynamics of the cl + ch4 -> hcl + ch3 and h + ch3cl reactions.
\newblock {\em J.~Chem.\ Phys. \bf 136\/} (2012), 044307.

\bibitem{li12:041103}
{\rm J.~Li, Y.~Wang, B.~Jiang, J.~Ma, R.~Dawes, D.~Xie, J.~M. Bowman,  \biband~
  H.~Guo}.
\newblock Communication: A chemically accurate global potential energy surface
  for the ho + co $\to$ h + co2 reaction.
\newblock {\em J.~Chem.\ Phys. \bf 136}, 4 (2012), 041103.

\bibitem{sha16:071101}
{\rm K.~Shao, J.~Chen, Z.~Zhao,  \biband~ D.~H. Zhang}.
\newblock Fitting potential energy surfaces with fundamental invariant neural
  network.
\newblock {\em J.~Chem.\ Phys. \bf 145\/} (2016), 071101.

\bibitem{pin18:26315}
{\rm L.~Ping, Y.~Zhu, A.~Li, H.~Song, Y.~Li,  \biband~ M.~Yang}.
\newblock Dynamics and kinetics of the reaction {OH + H$_2$S $\to$ H$_2$O + SH}
  on an accurate potential energy surface.
\newblock {\em Phys.\ Chem.\ Chem.\ Phys. \bf 20\/} (2018), 26315--26324.

\bibitem{lux18:174303}
{\rm X.~Lu, Q.~Meng, X.~Wang, B.~Fu,  \biband~ D.~H. Zhang}.
\newblock Rate coefficients of the h + h${_2}$o${_2}$ ->h${_2}$ + ho${_2}$
  reaction on an accurate fundamental invariant-neural network potential energy
  surface.
\newblock {\em J.~Chem.\ Phys. \bf 149\/} (2018), 174303.

\bibitem{chr20:104105}
{\rm C.~Schran, K.~Brezina,  \biband~ O.~Marsalek}.
\newblock Committee neural network potentials control generalization errors and
  enable active learning.
\newblock {\em J.~Chem.\ Phys. \bf 153\/} (2020), 104105.

\bibitem{lin20:154104}
{\rm Q.~Lin, Y.~Zhang, B.~Zhao,  \biband~ B.~Jiang}.
\newblock Automatically growing global reactive neural network potential energy
  surfaces: A trajectory-free active learning strategy.
\newblock {\em J.~Chem.\ Phys. \bf 152\/} (2020), 154104.

\bibitem{lin21:2691}
{\rm Q.~Lin, L.~Zhang, Y.~Zhang,  \biband~ B.~Jiang}.
\newblock Searching configurations in uncertainty space: Active learning of
  high-dimensional neural network reactive potentials.
\newblock {\em J.~Chem.\ Theory Comput. \bf 17\/} (2021), 2691--2701.

\bibitem{ele18:17414}
{\rm E.~Uteva, R.~S. Graham, R.~D. Wilkinson,  \biband~ R.~J. Wheatley}.
\newblock Active learning in gaussian process interpolation of potential energy
  surfaces.
\newblock {\em J.~Chem.\ Phys. \bf 149\/} (2018), 174114.

\bibitem{kol57:953}
{\rm A.~N. Kolmogorov}.
\newblock On the representation of continuous functions of many variables by
  superposition of continuous functions of one variable and addition.
\newblock {\em Dokl. Akad. Nauk SSSR \bf 114\/} (1957), 953--956.

\bibitem{liu25:041051}
{\rm Z.~Liu, M.~Tegmark, P.~Ma, W.~Matusik,  \biband~ Y.~Wang}.
\newblock Kolmogorov-arnold networks meet science.
\newblock {\em Phys. Rev. X \bf 15\/} (2025), 041051.

\bibitem{mey90:73}
{\rm H.-D. Meyer, U.~Manthe,  \biband~ L.~S. Cederbaum}.
\newblock The multi-configurational time-dependent {H}artree approach.
\newblock {\em Chem.\ Phys.\ Lett. \bf 165\/} (1990), 73--78.

\bibitem{man92:3199}
{\rm U.~Manthe, H.-D. Meyer,  \biband~ L.~S. Cederbaum}.
\newblock Wave-packet dynamics within the multiconfiguration {H}artree
  framework: General aspects and application to {NOCl}.
\newblock {\em J.~Chem.\ Phys. \bf 97\/} (1992), 3199--3213.

\bibitem{mey98:3011}
{\rm H.-D. Meyer}.
\newblock Multiconfiguration time-dependent {H}artree method.
\newblock In {\em {T}he {E}ncyclopedia of {C}omputational {C}hemistry\/}
  (Chichester, 1998), P.~v.~R.~Schleyer, N.~L. Allinger, T.~Clark,
  J.~Gasteiger, P.~A. Kollman, {H. F. {Schaefer III}}, \biband~ P.~R.
  Schreiner, Eds., vol.~5, John Wiley and Sons, pp.~3011--3018.

\bibitem{bec00:1}
{\rm M.~H. Beck, A.~J{\"a}ckle, G.~A. Worth,  \biband~ H.-D. Meyer}.
\newblock The multi-configuration time-dependent {H}artree ({MCTDH}) method:
  {A} highly efficient algorithm for propagating wave packets.
\newblock {\em Phys.~Rep \bf 324\/} (2000), 1--105.

\bibitem{mey03:251}
{\rm H.-D. Meyer \biband~ G.~A. Worth}.
\newblock {Q}uantum molecular dynamics: {P}ropagating wavepackets and density
  operators using the multiconfiguration time-dependent {H}artree ({MCTDH})
  method.
\newblock {\em Theor.\ Chem.\ Acc. \bf 109\/} (2003), 251--267.

\bibitem{mey09:book}
{\rm H.-D. Meyer, F.~Gatti,  \biband~ G.~A. Worth}, Eds.
\newblock {\em {Multidimensional Quantum Dynamics: MCTDH Theory and
  Applications}}.
\newblock Wiley-VCH, Weinheim, 2009.

\bibitem{mey12:351}
{\rm H.-D. Meyer}.
\newblock Studying molecular quantum dynamics with the multiconfiguration
  time-dependent {H}artree method.
\newblock {\em WIREs:~Comput.~Mol.~Sci. \bf 2\/} (2012), 351--374.

\bibitem{suk01:2604}
{\rm S.~Sukiasyan \biband~ H.-D. Meyer}.
\newblock On the effect of initial rotation on reactivity. {A}
  multi-configuration time-dependent {H}artree ({MCTDH}) wave-packet
  propagation study on the {H}+{D}$_2$ and {D}+{H}$_2$ reactive scattering
  systems.
\newblock {\em J.~Phys.\ Chem.~A \bf 105\/} (2001), 2604--2611.

\end{thebibliography}

\end{document}